\documentclass[conference]{IEEEtran}
\IEEEoverridecommandlockouts
\usepackage[left=1.62cm,right=1.62cm,top=1.9cm]{geometry}
\usepackage{cite}
\usepackage{amsmath,amssymb,amsfonts}
\usepackage{algorithmic}
\usepackage{graphicx}
\usepackage{textcomp}
\usepackage{amsmath}

\usepackage{amsmath}
\usepackage{mathtools}
\usepackage{verbatim}
\usepackage{amsmath}
\usepackage{xcolor}

\usepackage[T1]{fontenc}
\usepackage[utf8]{inputenc}
\usepackage{authblk}

\def\BibTeX{{\rm B\kern-.05em{\sc i\kern-.025em b}\kern-.08em
    T\kern-.1667em\lower.7ex\hbox{E}\kern-.125emX}}
\begin{document}

\title{Active RIS-Aided Terahertz Communications with Phase Error and Beam Misalignment \\

{%\footnotesize \textsuperscript{*}Note: Sub-titles are not captured in Xplore and should not be used
} \thanks{This work was partly supported by “Regional Innovation Strategy (RIS)” through the National Research Foundation of Korea (NRF), funded by the Ministry of Education (MOE) (2021RIS-004). It was also partly supported by the Institute of Information \& communications Technology Planning \& Evaluation (IITP) grant funded by the Korea government (MSIT) (2022-0-00704, Development of 3D-NET Core Technology for High-Mobility Vehicular Service), and by the Basic Science Research Program through the NRF funded by the MOE (NRF-2022R1I1A1A01071807, 2021R1I1A3041887).}
}

\author[*]{Waqas Khalid}
\author[$\dagger$]{Heejung Yu}
\author[$\ddagger$]{Farman Ali}
\author[$\vert$]{Huiping Huang}
\affil[*]{Institute of Industrial Technology, Korea University, Sejong, Korea; waqas283@\{korea.ac.kr, gmail.com\}}
\affil[$\dagger$]{Dept. of Elec. \& Inform. Eng., Korea University, Sejong, Korea; heejungyu@korea.ac.kr}
\affil[$\ddagger$]{Dept. of Applied AI, School of Convergence, Sungkyunkwan University, Seoul, Korea; farman0977@skku.edu}
\affil[$\vert$]{Dept. of Electrical Engineering, Chalmers University of Technology, Sweden; huiping@chalmers.se}
\renewcommand\Authands{ and }

%\author{\IEEEauthorblockN{Waqas Khalid}
%\IEEEauthorblockA{\textit{Institute of %Industrial Technology} \\
%\textit{Korea University}\\
%Sejong 30019, South Korea \\
%waqas283@korea.ac.kr; waqas283@gmail.com}
%\and

%\IEEEauthorblockN{M. Atif Ur Rehman}
%\IEEEauthorblockA{\textit{Dept. of %Computing \& Mathematics}\\
%Manchester Metropolitan University\\
%Manchester M1 5GD, UK\\
%m.atif.ur.rehman@mmu.ac.uk}
%\and
%\IEEEauthorblockN{Heejung Yu}
%\IEEEauthorblockA{\textit{Dept. of Elec. \& %Inform. Eng.} \\
%\textit{Korea University}\\
%Sejong 30019, South Korea \\
%heejungyu@korea.ac.kr}

%}

\maketitle

\begin{abstract}
Terahertz (THz) communications will be pivotal in sixth-generation (6G) wireless networks, offering significantly wider bandwidths and higher data rates. However, the unique propagation characteristics of the THz frequency band, such as high path loss and sensitivity to blockages, pose substantial challenges. Reconfigurable intelligent surfaces (RISs) present a promising solution for enhancing THz communications by dynamically shaping the propagation environment to address these issues. Active RISs, in particular, can amplify reflected signals, effectively mitigating the multiplicative fading effects in RIS-aided links. Given the highly directional nature of THz signals, beam misalignment is a significant concern, while discrete phase shifting is more practical for real-world RIS deployment compared to continuous adjustments. This paper investigates the performance of active-RIS-aided THz communication systems, focusing on discrete phase shifts and beam misalignment. An expression for the ergodic capacity is derived, incorporating critical system parameters to assess performance. Numerical results offer insights into optimizing active-RIS-aided THz communication systems. 
\end{abstract}


\section{Introduction}
Terahertz (THz) communications are being considered for future sixth-generation (6G) wireless networks due to their potential to provide extremely high bandwidth and data rates, enabling ultra-fast communication speeds \cite{V0}. These capabilities are crucial for meeting the substantial data demands of next-generation applications, such as holographic communications, ultra-high-definition video streaming, and extensive Internet of Things (IoT) deployments. The  stringent requirements of 6G networks, including significantly higher data rates, ultra lower latency, and enhanced connectivity make THz frequencies a promising solution for supporting these advanced features.

Reconfigurable intelligent surfaces (RISs) are being considered for 6G networks due to their ability to manipulate electromagnetic (EM) waves. By reflecting and refracting signals, RISs can enhance signal propagation, mitigate issues such as blockage and path loss, and improve coverage. RISs also enable adaptive beamforming and spatial multiplexing, boosting spectral efficiency. They enhance the energy efficiency of 6G networks due to their low power consumption, reducing the overall energy footprint while maintaining high-performance communication \cite{V1}.

RISs are particularly suitable for THz communications because they address the highly directional nature and propagation challenges of the THz frequency band \cite{V2}. However, passive RISs face performance limitations due to double fading across RIS-cascaded links. Active RISs can amplify reflected signals, mitigating the multiplicative fading effects that passive RISs cannot. By incorporating active components, active RISs adjust both the phase and amplitude of signals, significantly improving the signal-to-noise ratio (SNR). This results in enhanced coverage and more reliable communication, especially in environments with high path loss or complex propagation conditions \cite{V3}.

\subsection{Motivation and Contribution}

Beam misalignment poses a major challenge in THz communications due to the highly directional nature of THz signals, which demand precise alignment between the transmitter and receiver to ensure a stable connection \cite{V16}. The reflection phases of RIS elements are often assumed to be precisely adjustable, allowing continuous phase shifts within the range of $[0, 2\pi)$. However, this level of precision is impractical due to inherent hardware constraints, as each RIS element can typically achieve only a limited set of discrete phase shifts. 

The combined impact of beam misalignment and discrete phase shifts on system performance has not been thoroughly investigated, which is the primary focus of this paper. To accurately predict signal behavior and optimize system design, it is essential to consider factors such as molecular absorption, beam misalignment, signal amplification, active noise, and quantization errors. This paper examines active RIS-aided THz communications by deriving the ergodic capacity while incorporating these considerations. Through mathematical analysis and numerical simulations, we provide insights into capacity degradation caused by quantization errors and beam misalignment, as well as performance improvements achieved with active RIS.

\begin{figure}[t]
\centering
\includegraphics[width=3.2in,height=3in]{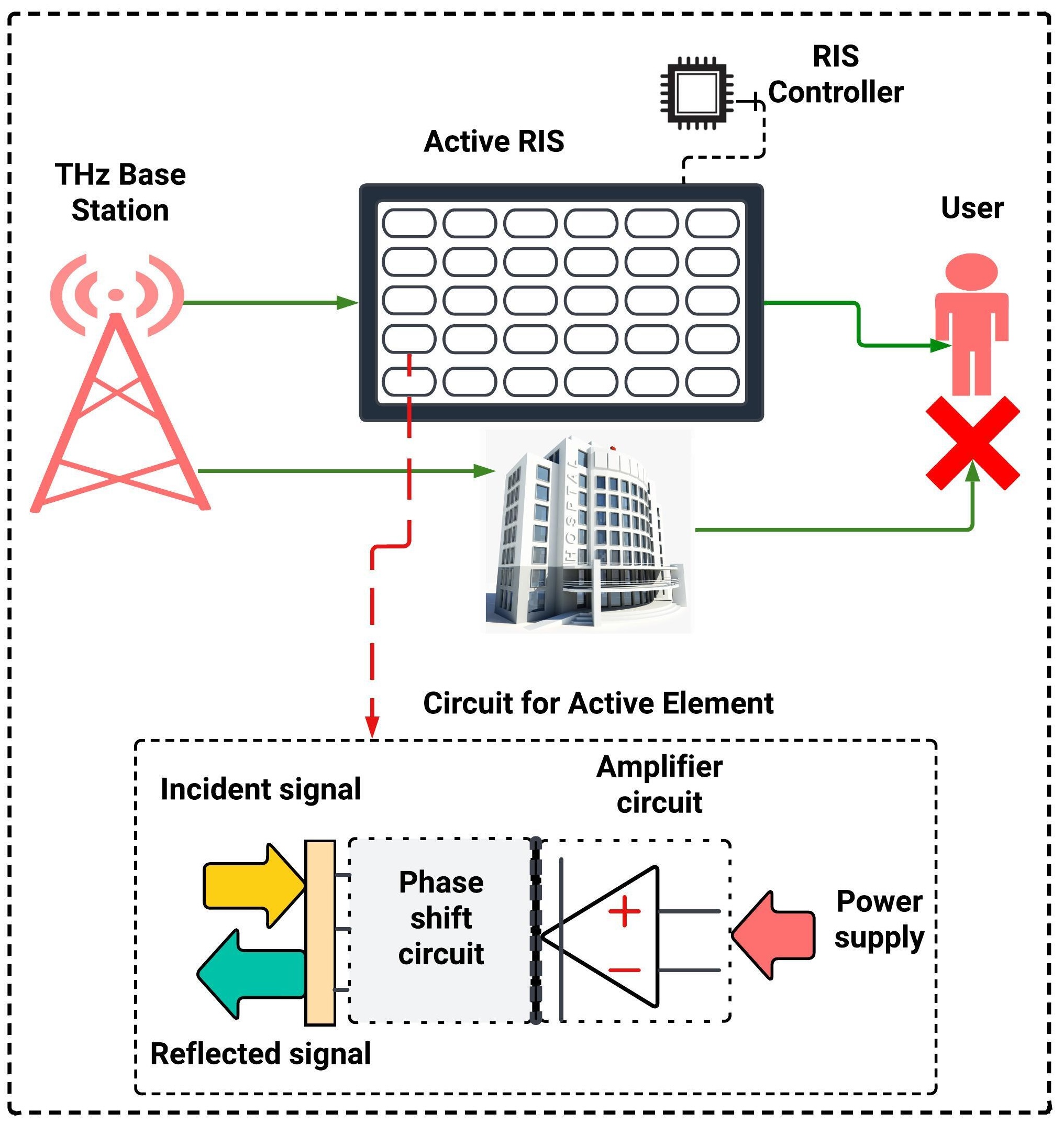}
\caption {System model.}
\label{diagram1}
\end{figure}

\section{System Model}

We consider a THz communication scenario as depicted in Fig. 1, where the direct path between the base station (BS) and the user ($U$) is obstructed. To overcome this blockage, an active RIS with $M$ elements is deployed to facilitate communication through RIS-cascaded links. Each element of the active RIS is equipped with power amplifiers and phase-shift circuits, allowing for both signal reflection and amplification. Due to the significant path loss in THz communications, further exacerbated by molecular absorption and beam misalignment, we assume that the signal power from multiple reflections within the RIS is negligible. The channel fading coefficients between the BS and RIS ($f_m$) and between the RIS and the user ($g_m$) are modeled as complex Gaussian random variables. The channel envelopes are assumed to be independently Rayleigh distributed, with $f_m \sim \mathcal{CN}(0,1)$ and $g_m \sim \mathcal{CN}(0,1)$ \cite{V15a,V9}.

\subsection{Path Gain Coefficient}
The path gain coefficient, which characterizes the overall signal attenuation between the base station (BS) and the user $U$ through the active RIS, is given by $h_L = h_P h_{A}$. Here, $h_P$ represents the propagation gain, and $h_{A}$ denotes the molecular absorption gain. The propagation gain, $h_P$, is modeled using the Friis transmission equation as $h_P = \frac{c\sqrt{G_1 G_2}}{8\sqrt{\pi^3}fd_1 d_2}$, where $G_1$ and $G_2$ are the antenna gains of the BS and user, dependent on their orientations, $c$ is the speed of light, $f$ is the operating frequency, and $d_1$ and $d_2$ are the distances from the BS to the RIS and from the RIS to the user, respectively. 

The molecular absorption gain, $h_{A}$, is expressed as $h_{A} = \exp{\left(-\frac{\kappa(f)(d_1 + d_2)}{2}\right)}$, where $\kappa(f)$ is the frequency-dependent molecular absorption coefficient \cite{V16}. Determining the path gain coefficient $h_{L}$ is crucial for optimizing the performance of RIS-aided THz communication systems, especially in scenarios where the direct communication path is obstructed. This detailed modeling enables more accurate simulations and performance evaluations, which are essential for the effective design and deployment of active RIS-aided THz networks.

\subsection{Beam Misalignment Coefficient}

Beam misalignment poses a significant challenge in THz communications due to the highly directional nature of THz beams, which require precise alignment between the base station (BS) and user ($U$) to ensure effective signal transmission \cite{V16}. The beam misalignment coefficient, $h_{M}$, is characterized by its probability density function (PDF):

\[
f_{h_{M}}(x) = \zeta \phi^{-\zeta} x^{\zeta-1}, \quad 0 \leq x \leq \phi,
\]
where $\phi = (\operatorname{erf}(l))^2$ represents the power fraction captured by the user under perfect alignment, with $\operatorname{erf}(.)$ denoting the error function. The term $l = \frac{\sqrt{\pi}a}{\sqrt{2}\omega_{BS}}$ is defined by the radius $a$ of the user’s effective area and the BS beam footprint $\omega_{BS}$. The parameter $\zeta = \frac{\omega_{e}^2}{4\sigma_s^2}$ is determined by the equivalent beam width $\omega_{e}$ and the variance of the beam misalignment displacement $\sigma_s^2$.

This power-law distribution models the probability of beam misalignments in THz communications, capturing the impact of beam width and displacement variance on received signal power. As beam misalignment can reduce signal power and degrade system performance, modeling the beam misalignment coefficient is necessary for optimizing THz systems. Incorporating these factors into performance analysis is crucial for evaluating link robustness and addressing challenges from narrow beam widths and the high sensitivity of THz frequencies.

\section{Performance Analysis}

The received signal at $U$ can be expressed as follows:

\begin{align} \label{eq4}
y=&\underbrace{\sqrt{P_s}h_{L}h_{M}  \beta \left(\sum^{M}_{m=1}f_mg_m e{^{j\theta_m}}\right)x}_{\text{desired signal}}
\nonumber \\
&+\underbrace{\beta \left(\sum^{M}_{m=1}g_m e{^{j\theta_m}}\right)n_r}_{\text{noise due to active RIS}}+n_u
\end{align}
where $\theta_m$ is the induced phase for the $m$th element, $\beta$ is the amplification gain (which can exceed one for the active RIS), $x$ is the transmitted signal from the BS with $\mathbb{E}[|x|^2] = 1$, $P_s$ is the transmission power, $n_r \sim \mathcal{CN} (0,\sigma_r^2)$ is the noise at the active RIS, and $n_u \sim \mathcal{CN} (0,\sigma_u^2)$ is the noise at $U$. 

It is assumed that the instantaneous channel state information (CSI) is available at the BS, allowing the phases of $f_m$ and $g_m$ to be known at the active RIS \cite{V5,V6}. However, the phase shifts applied at the active RIS are discrete, selected from the set $\mathcal{F}=\left\{0,\frac{2\pi}{2^b},...,\frac{\left(2^b-1\right)2\pi}{2^b}\right\}$, where $b$ is the number of quantization bits. Due to this quantization of phase shifts, the phases cannot be perfectly aligned, leading to phase shift quantization errors. These errors are defined as $\Phi_m=\theta_m-\arg(f_m)-\arg(g_m)$ and are uniformly distributed within the interval $\left[-\frac{\pi}{2^b},\frac{\pi}{2^b}\right]$ \cite{V18}.

The ergodic capacity of $U$ can be expressed as, 

\begin{align} \label{eq5}
C=\frac{1}{\ln2}\int_0^\infty\frac{1-F_\gamma(s)}{1+s}ds
\end{align}
where $\gamma$ is the SNR at $U$, which can be expressed as,

\begin{align} \label{eq7}
\gamma= \rho_s \beta^2 |h_{M}| ^2\;\left(\sum^{M}_{m=1}|f_m| |g_m|e^{j\Phi_m}\right)^2
\end{align}
where $\rho_s=\frac{P_s h_{L}^2}{\beta^2\sum^{M}_{m=1}|g_m|^2\sigma^2_{r}+\sigma^2_{u}}$.

The closed-form expression for the ergodic capacity of \( U \) can be derived by transforming Eq. 2 using the Meijer G-function, which facilitates the computation of complex integrals and helps evaluate system performance under various conditions. This analysis examines the effects of key parameters, including amplification gain, RIS element count, quantization-induced phase errors, beam misalignment, and active noise.

\begin{comment}This approach provides an in-depth examination of the interactions between these factors and their impact on the efficiency and reliability of active RIS-aided THz communication systems.\end{comment}

\section{Numerical Results}

To validate the derived expression, we provide numerical results using the following default parameters: antenna gains $G_1 = G_2 = 30$ dBi, distances $d_1 = d_2 = 15$ m, operating frequency $f = 0.3$ THz, number of RIS elements $M = 100$, amplification factor $\beta = 2$, noise variances $\sigma_r^2 = 0.01$ (at the RIS) and $\sigma_u^2 = 0.05$ (at the user $U$), molecular absorption factor $v = 0.68$, beam misalignment parameters $\phi = 0.2$ and $\zeta = 0.52$, BS transmit power $P_s \in {-20, 30}$ dB, and the number of quantization bits $b = 2$. 

Fig. 2 illustrates the ergodic capacity $C$ as a function of $\rho_s$, showing the effects of quantization-induced phase errors and beam misalignment. The results demonstrate that both factors significantly degrade the ergodic capacity, with their impact becoming more pronounced as $\rho_s$ increases. This underscores the importance of precise phase control and accurate alignment in active-RIS-aided THz communication systems.

\begin{figure}[t]
\centering
\includegraphics[width=3in,height=3.2in]{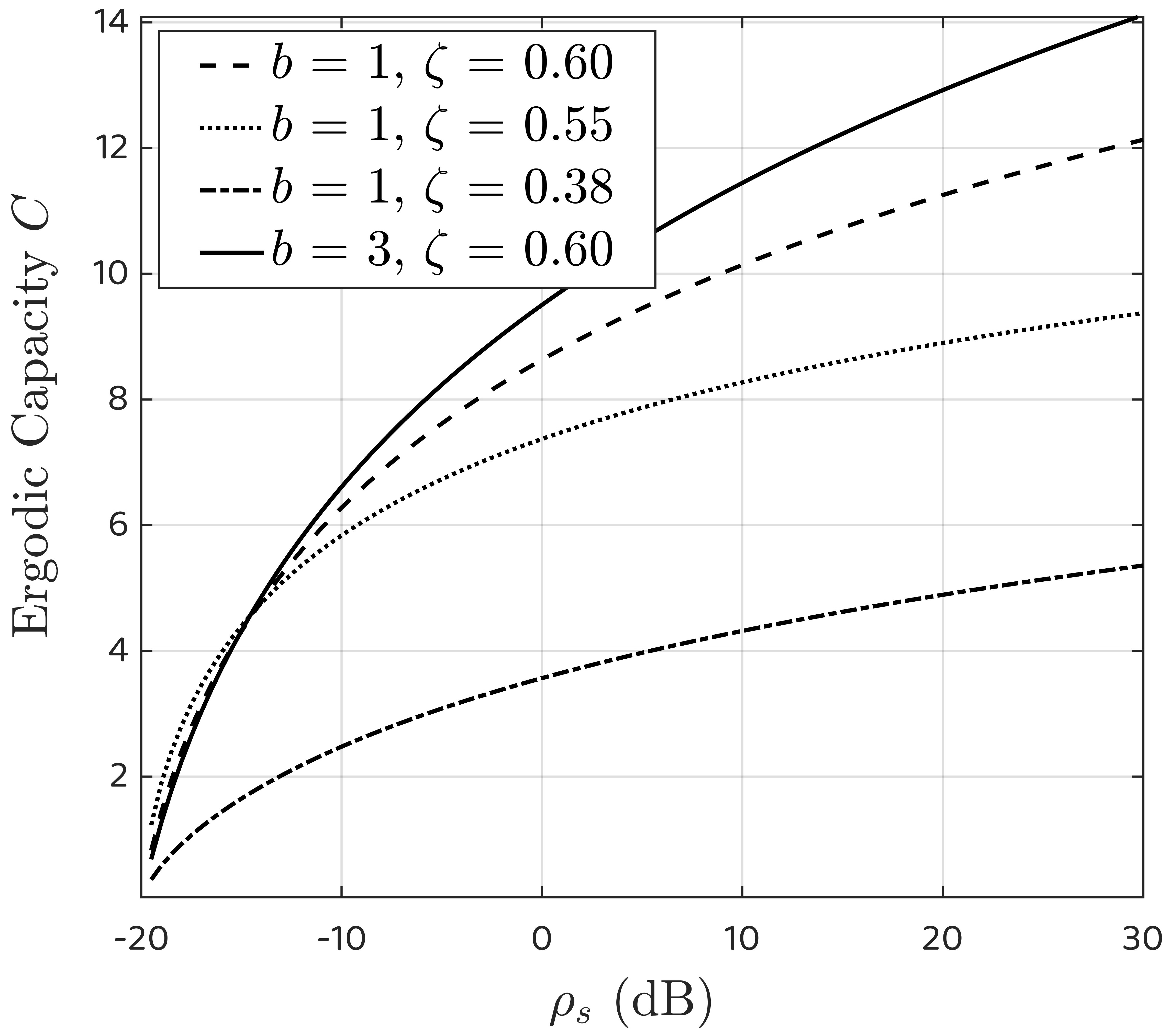}
\caption {Ergodic capacity ($C$) vs. SNR ($\rho_s$)}
\label{diagram1}
\end{figure}

\section{Conclusion}

In this paper, we analyzed the performance of active-RIS-aided THz communication systems by formulating the ergodic capacity while accounting for key factors such as discrete phase shifts, beam misalignment, molecular absorption, and active noise. Our study offered a preliminary analysis of these factors and their impact on system performance, laying the groundwork for more detailed future research. \begin{comment}Our numerical results demonstrated the significant impact of quantization-induced phase errors and beam misalignment on ergodic capacity.\end{comment}

%
%\section*{Acknowledgment}

%Conceptualization, W.Khalid, H. Yu,; writing—original-draft preparation, writing—review and editing. All
%authors have read and agreed to the published version of the manuscript.

%\section*{References}

\end{document}